\documentclass[10pt,journal]{IEEEtran}
\usepackage{array}
\usepackage{graphicx}
\usepackage{epsfig}
\usepackage{cite}
\usepackage{epstopdf}
\usepackage{amsfonts,amsmath,amsthm,amssymb,balance}
\usepackage{multirow,balance}
\usepackage{bm}
\usepackage[usenames,dvipsnames]{color}
\usepackage{colortbl}
\usepackage{float}
\usepackage{subfigure}
\usepackage{tabularx}
\usepackage{enumerate}
\usepackage{algorithm}
\usepackage{algorithmic}
\usepackage[marginal]{footmisc}
\usepackage{url}
\usepackage{caption}
\usepackage{color}
\usepackage{soul}

\begin{document}

\title{Multi-layer RIS on Edge: Communication, Computation and Wireless Power Transfer}

\author{ Shuyi Chen,~\IEEEmembership{Member,~IEEE}, Junhong Jia, Baoqing Zhang, Yingzhe Hui, Yifan Qin, Weixiao Meng,~\IEEEmembership{Senior Member,~IEEE}, Tianheng Xu,~\IEEEmembership{Member,~IEEE}
\thanks{ 
~~Shuyi Chen (e-mail: {\tt chenshuyitina@gmail.com}), Junhong Jia (e-mail:{\tt 24s005051@stu.hit.edu.cn}), Yingzhe Hui (e-mail: {\tt yingzhe\_hui@stu.hit.edu.cn}) and Weixiao Meng (e-mail: {\tt wxmeng@hit.edu.cn}) are with the Communication Research Center and the China-Chile ICT Belt and Road Joint Laboratory, Harbin Institute of Technology, China. Baoqing Zhang (e-mail: {\tt bqzhang10@163.com}) is with Beijing Institute of Electronic System Engineering, China. Yifan Qin (e-mail: {\tt qinyifan@hrbeu.edu.cn}) is with the Ministry of Education Key Lab of In-fiber Integrated Optics, Harbin Engineering University, China. Tianheng Xu (e-mail: {\tt xuth@sari.ac.cn}) is with the Shanghai Advanced Research Institute, Chinese Academy of Sciences, China.}
\thanks{
~~This work was supported in part by the China’s key R \& D Program No. 2020YFE0205800, the National Science Fund for Young Scholars No. 62201176,  Natural Science Foundation of Heilongjiang Province of China No. YQ2023005,   Young Elite Scientist Sponsorship Program by CAST No. YESS20210339.}
\thanks{
~~The paper was submitted \today}
}

\date{\today}
\markboth{IEEE Internet of Things Magazine, Vol. XX, No. YY, Month 2024} {Shell \MakeLowercase{\textit{et al.}}: Bare Demo of IEEEtran.cls for IEEE Journals}
	
\renewcommand{\baselinestretch}{1.2}
	
\maketitle

\begin{abstract}
The rapid expansion of Internet of Things (IoT) and its integration into various applications highlight the need for advanced communication, computation, and energy transfer techniques. However, the traditional hardware-based evolution of communication systems faces challenges due to excessive power consumption and prohibitive hardware cost. With the rapid advancement of reconfigurable intelligent surface (RIS), a new approach by parallel stacking a series of RIS, i.e., multi-layer RIS, has been proposed. Benefiting from the characteristics of scalability, passivity, low cost, and enhanced computation capability, multi-layer RIS is a promising technology for future massive IoT scenarios. Thus, this article proposes a multi-layer RIS-based universal paradigm at the network edge, enabling three functions, i.e., multiple-input multiple-output (MIMO) communication, computation, and wireless power transfer (WPT). Starting by picturing the possible applications of multi-layer RIS, we explore the potential signal transmission links, energy transmission links, and computation processes in IoT scenarios, showing its ability to handle on-edge IoT tasks and associated green challenges. Then, these three key functions are analyzed respectively in detail, showing the advantages of the proposed scheme, compared with the traditional hardware-based scheme. To facilitate the implementation of this new paradigm into reality, we list the dominant future research directions at last, such as inter-layer channel modeling, resource allocation and scheduling, channel estimation, and edge training. It is anticipated that multi-layer RIS will contribute to more energy-efficient wireless networks in the future by introducing a revolutionary paradigm shift to an all-wave-based approach.

\end{abstract}

\IEEEpeerreviewmaketitle

\vspace{0.4in}
\section{Introduction}
5G  mobile communication systems have entered the stage of commercial deployment and gradually permeated into all facets of society. Although the 5G network is a key enabler for IoT, it is unlikely to meet the evolving demands of emerging applications and performance requirements. Building upon 5G, 6G will not only continue to enhance IoT capabilities but will also shift its focus from rate-centric services to ultra-low latency and energy-efficient communications. This transition presents a significant challenge due to the rapid growth in the number of connected devices and the ever-increasing volume of network traffic.

In the era of IoT,  billions of sensors/devices will be ubiquitously connected. IoT embodies the communication and interaction between objects and embedded devices from diverse environments to collect raw data. The data is then analyzed to extract valuable information, which is used to drive  decision-making and automate processes. By providing crucial services such as information exchange and monitoring, IoT opens up limitless potential applications. As it continues to evolve, IoT plays an increasingly significant role in every aspect of life, enabling a more intelligent and efficient world.

However, there are still issues to be solved with IoT devices. Due to size constraints and diverse usage scenarios, IoT devices are typically equipped with quite limited energy supply. Currently, batteries remain to be the most prevalent solution for powering electronics. ‌Nevertheless, frequent battery replacements and recharging plugins are cost-prohibitive and potentially hazardous in extreme and inaccessible conditions. Therefore, it is imperative to eliminate the reliance on batteries in IoT devices and delve into alternative energy sources for facilitating the realization of self-powered devices.

On the other hand,  the increasing number of IoT devices implies that vast amounts of data will be generated rapidly, necessitating efficient transmission and processing. With the extensive applications of IoT, the total installed IoT-based devices are projected to amount to approximately 41.6 billion, and nearly 79.4 Zettabytes of data may be generated and consumed in 2025~\cite{numIOT}. Therefore, the implementation of MIMO techniques at BS is essential to accommodate the large number of IoT devices. However, further scaling of antenna arrays has become the bottleneck in the evolution of traditional communication networks, mainly due to excessive energy consumption and prohibitive hardware cost. What’s more, network congestion is a critical issue when offloading such huge data from edge to cloud for data processing. Therefore, a more applicable way is to perform data processing in proximity to data sources, which gives birth to a brand-new computation paradigm, edge computing. 

Artificial intelligence (AI) is a tool that can effectively and efficiently mine valuable information from vast volumes of data and make decisions. The marriage of edge computing, IoT, and AI is natural given their inherent complementarity. However, state-of-the-art deep learning-based AI systems demand tremendous computation and communication resources, resulting in not only substantial latency and energy consumption but also severe network congestion. This situation necessitates an update in hardware paradigm due to the limited computational power of edge servers and IoT devices.

Meanwhile, RIS has seen exponential growth in recent years. Due to its exceptional capability to dynamically manipulate incident electromagnetic (EM) waveforms, RIS enables numerous wireless communication applications while reducing cost, size, and power consumption, thereby driving a paradigm shift toward energy-efficient communication systems. However, a single-layer nearly passive RIS architecture fails to implement advanced signal processing functionalities. Therefore, a multi-layer RIS architecture can be established by cascading a series of RISs, which has a similar structure to artificial neural networks. This architecture allows for the execution of various complex signal processing and computation tasks as EM waves propagate through its multiple layers.

In summary, IoT devices continue to face green challenges in MIMO configuration, AI implementation, and energy supply. A key issue worth investigating is how to address these challenges with minimal addition of new modules, keeping hardware cost and power consumption as low as possible, while ideally enabling easy upgrades to existing IoT devices and communication infrastructure. Inspired by recent advancements in multi-layer RIS-based prototypes and theoretical research, we aim to explore the applications of multi-layer RIS in edge IoT scenarios. In different configuration modes, multi-layer RIS can perform MIMO transmission, computation, and WPT while transmitting EM waves through its RIS layers. Compared to traditional digital hardware methods, the multi-layer RIS-based approaches demonstrate potential in scalability with low cost, enabling light-speed, power-free computation, and achieving more precise power transfer. It is worth noting that research on multi-layer RIS is still in its early stages. Applying multi-layer RIS in edge IoT scenarios is an unprecedented attempt, as it has the potential to address these challenges with a universal paradigm.

In this article, we propose a multi-layer RIS-based universal paradigm to facilitate communication, computation, and long-term operation of IoT devices on edge.  The rest of the article can be outlined as follows. First, we briefly introduce the basic knowledge of multi-layer RIS and the principles of RIS-based communication, computation, and WPT. Next, we present a system model that outlines potential links and the functionalities of multi-layer RIS at the network edge. The advantages and applications of multi-layer RIS-based MIMO, computation, and WPT in edge IoT scenarios are then discussed respectively, followed by future directions and conclusions.

\section{Fundamentals of Multi-layer RIS }
\subsection{Multi-layer RIS structure}

\begin{figure*}
\centering
\includegraphics[width=15cm]{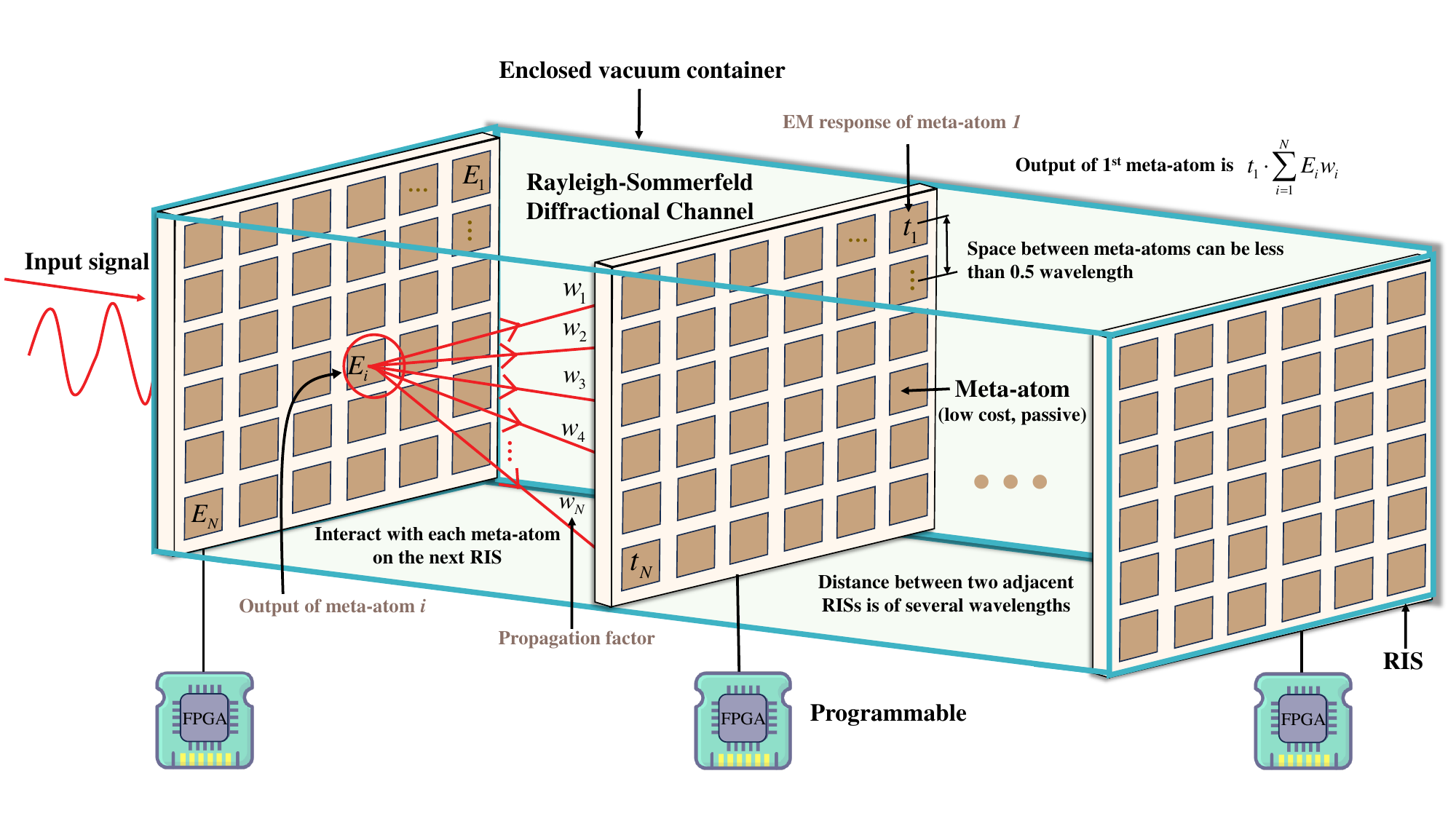}
\captionsetup{font={small}}
\caption{The structure of multi-layer RIS. The entire structure should be enclosed within a container lined with absorbing materials to minimize energy loss and shield the inter-layer channels from external interference. The size of each meta-atom typically ranges from 1/10 to 1/2 of the wavelength. Meanwhile, the spacing between adjacent RIS layers is generally on the same order of magnitude as the wavelength. The distance between adjacent layers and the size of each meta-atom can be adjusted to meet the requirements of specific tasks, serving as optimizable parameters. Owing to the compact inter-layer spacing and the dense arrangement of passive meta-atoms, multi-layer RIS achieves a miniaturized structure with high scalability.}
\label{RIS}

\end{figure*}

As shown in Fig.~\ref{RIS}, multi-layer RIS is a compact enclosed vacuum container composed of several parallel stacked transmissive RIS. Each RIS is a programmable metasurface consisting of a large number of meta-atoms, arranged in a specific array, with adjustable EM properties. Each programmable meta-atom is passive and low-cost, typically consisting of a metal pattern, dielectric, and a tunable component. By adjusting the bias voltage of the meta-atom through a field-programmable gate array (FPGA), its EM response (phase, amplitude, and polarization, etc.) can be modified, enabling precise manipulation over the EM behavior of the penetrating wave. Each meta-atom can be analogized to a neuron, and the entire multi-layer RIS architecture can be viewed as the physical realization of a deep neural network. Thus, improved signal processing capability and computational power can be achieved by multi-layer RIS compared to its single-layer counterpart. This capability allows for the execution of complex, customized functions such as MIMO transmissions, computation, and WPT.

It is important to introduce the EM propagation mechanism between layers of multi-layer RIS. Based on the Huygens-Fresnel principle, the EM wave passing through each meta-atom on the previous RIS layer acts as a point source, illuminating all the meta-atoms on the subsequent RIS layer. In other words, the output of a meta-atom on the second layer is determined by the product of its EM response and the weighted sum of the outputs from all meta-atoms on the previous layer. The weights correspond to the inter-layer channel propagation factors, which depend on parameters such as the wavelength, distance and angle between the two meta-atoms on adjacent layers. These factors can be described using the Rayleigh-Sommerfeld diffraction theory.

\subsection{Principles of RIS-based communication, computation, and WPT}

\subsubsection{RIS-based MIMO communication}
\begin{figure*}
\centering
\includegraphics[width=16cm]{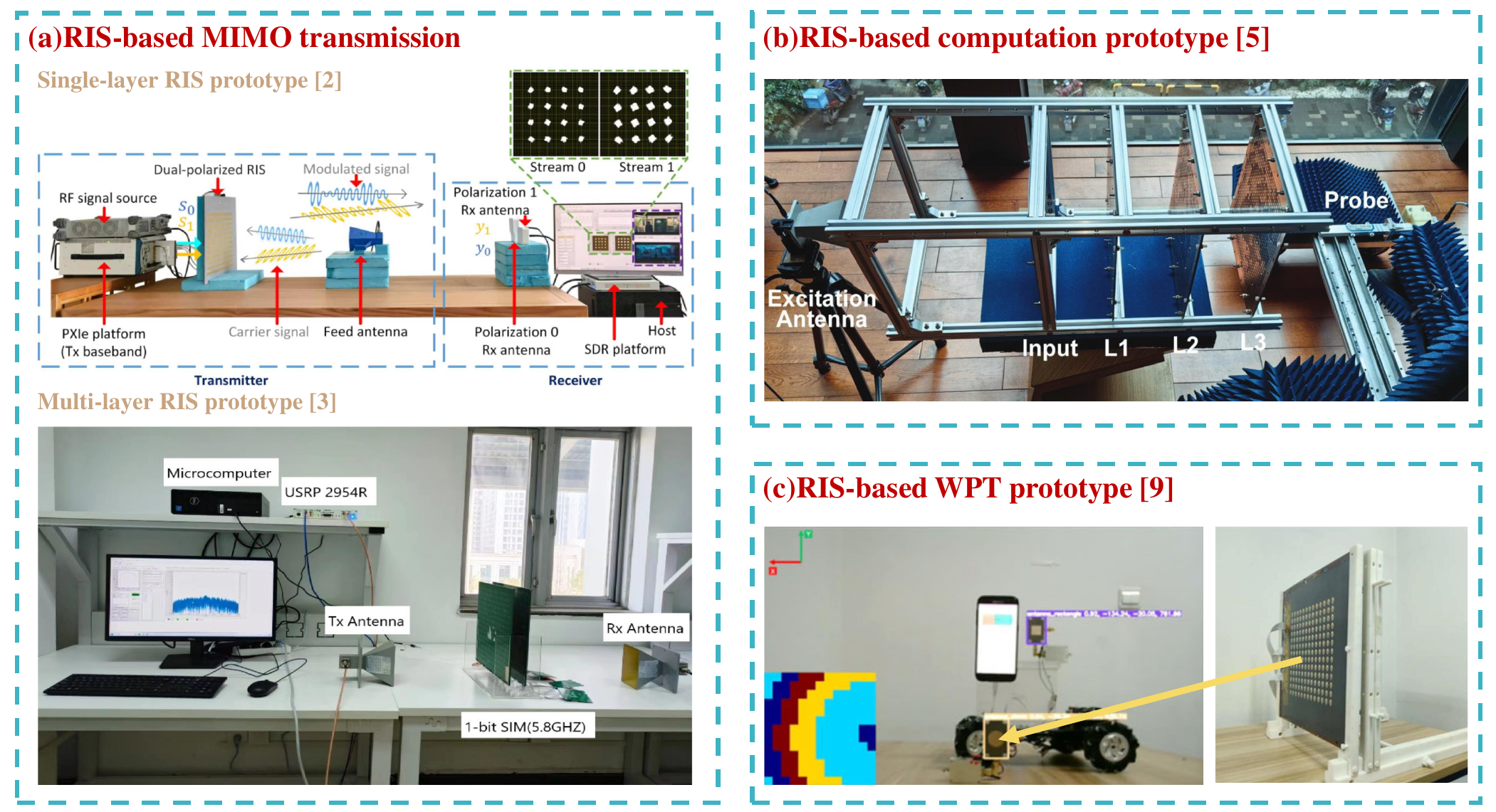}
\captionsetup{font={small}}
\caption{Prototypes of RIS-based MIMO transmission, computation and WPT, where, (a) shows two prototypes of RIS-based MIMO transmission. The first is a single-layer RIS prototype designed for point-to-point MIMO video transmission. The second is a multi-layer RIS prototype, which demonstrates enhanced performance compared to the single-layer prototype when the inter-spacing between adjacent layers is less than half the wavelength. (b) shows a three-layer RIS-based computation prototype that successfully performs handwritten digit classification at X-band, achieving a processing latency of less than 10 ns. (c) shows a RIS-based prototype realizing WPT and SWIPT on the right. The dimensions of the prototype are around 0.5m$\times$0.5m$\times$0.04m, close to the size of a regular household appliance.}

\label{prototype}

\end{figure*}

Given that the response of each meta-atom, including its phase and amplitude, can be adjusted, a RIS can function as a transmit antenna with reconfigurable EM properties. This capability enables a cost-effective and energy-efficient implementation of MIMO systems. A prototype of a RIS-based MIMO system~\cite{MIMOprototype1} is depicted in Fig.~\ref{prototype}, where a dual-polarized RIS-based 2$\times$2 MIMO transmission system is established, capable of transmitting two video streams at a data rate of 20 Mbps. Specifically, modulation is achieved by adjusting the meta-atoms in real time according to the data intended for transmission. As the single-tone carrier emitted by the RF signal source passes through the RIS, information is directly modulated onto the carrier. If all the meta-atoms of the RIS are controlled by the same external signal, single-input single-output communication is realized, as the same modulation is applied to the carrier signal. However, if each meta-atom can be independently controlled, an additional beamforming matrix can be applied to the carrier signal as it passes through the RIS during the modulation process, enabling MIMO transmissions. Additionally, a multi-layer RIS-based communication prototype is presented in~\cite{MIMOprototype2}, with experimental evaluations demonstrating its enhanced performance.

\subsubsection{RIS-based computation}
Many works suggest the use of physical layer solutions with over-the-air or all-wave-based neural networks to enable efficient computation, known as diffractional deep neural networks ($\rm D^2NN$)~\cite{semantic}. Multi-layer RIS is a promising candidate for implementing reconfigurable $\rm D^2NN$, which has the potential to serve as a physical layer computation device. Certain demonstrations of multi-layer RIS-based $\rm D^2NN$ are elaborated below.
\begin{itemize}
\item\textbf{Fully-connected layers:} 
  Since signal diffraction occurs between any two adjustable meta-atoms on adjacent layers according to Rayleigh-Sommerfeld diffraction theory, a multi-layer RIS structure can be interpreted as fully-connected layers. A three-layer $\rm D^2NN$ prototype with fully-connected layers~\cite{classification}, as shown in Fig.~\ref{prototype}, successfully implemented handwritten digit classification in the microwave frequency.
 \item\textbf{Shortcut connections:} 
 Shortcut connections can be realized by a RIS with the capability to regulate the EM properties of different polarization channels independently. The input signal is initially encoded concurrently in two polarization channels. As the signal propagates through multi-layer RIS, one channel undergoes task-specific manipulation, while the other preserves the original input features through fixed phase and amplitude manipulation. The signals from two channels then perform coordinated calculations with the same weight, mimicking a shortcut connection that can be used to construct a ResNet~\cite{shortcut}.
\item\textbf{Convolution layers:} 
A diffraction-driven multi-kernel optical convolution unit was fabricated on a silicon-on-insulator platform in~\cite{conv} to enable on-chip parallel convolution processing. The architecture was further employed to build optical convolutional neural networks for classification tasks.

\item\textbf{Activation functions:} 
It was experimentally demonstrated in~\cite{nonlinear} that nonlinear activation functions can be realized by letting the amplifier work in the saturation zone, which is qualitatively similar to the positive half of the sigmoid function. This enables the establishment of a deeper $\rm D^2NN$ with improved computational power.
\end{itemize}

\subsubsection{RIS-based WPT}
To minimize loss during EM wave propagation, advanced WPT systems typically employ phased arrays for energy beamforming, directing beams toward targeted devices to enhance RF power transfer efficiency. However, the utilization of phased arrays usually results in bulkiness and complexity, limiting their capability for scalability and large-scale deployments. RIS offers a promising alternative to phased arrays in WPT, leveraging its inherent advantages of low power consumption, cost-effectiveness, and compact size. Fig.~\ref{prototype}~\cite{WPTprototype} illustrates a reconfigurable power router for a RIS-based WPT scheme, which includes a plane-wave feeder and a transmissive 2-bit RIS-based beam generator. With the  assistance of an intelligent computation unit, this system dynamically delivers wireless power to single or multiple power-consuming targets based on real-time information from the environment. Additionally, the prototype demonstrates its simultaneous wireless information and power transfer (SWIPT) capability by transmitting the author's school badge while simultaneously illuminating an LED.

\begin{figure*}
\centering
\includegraphics[width=16cm]{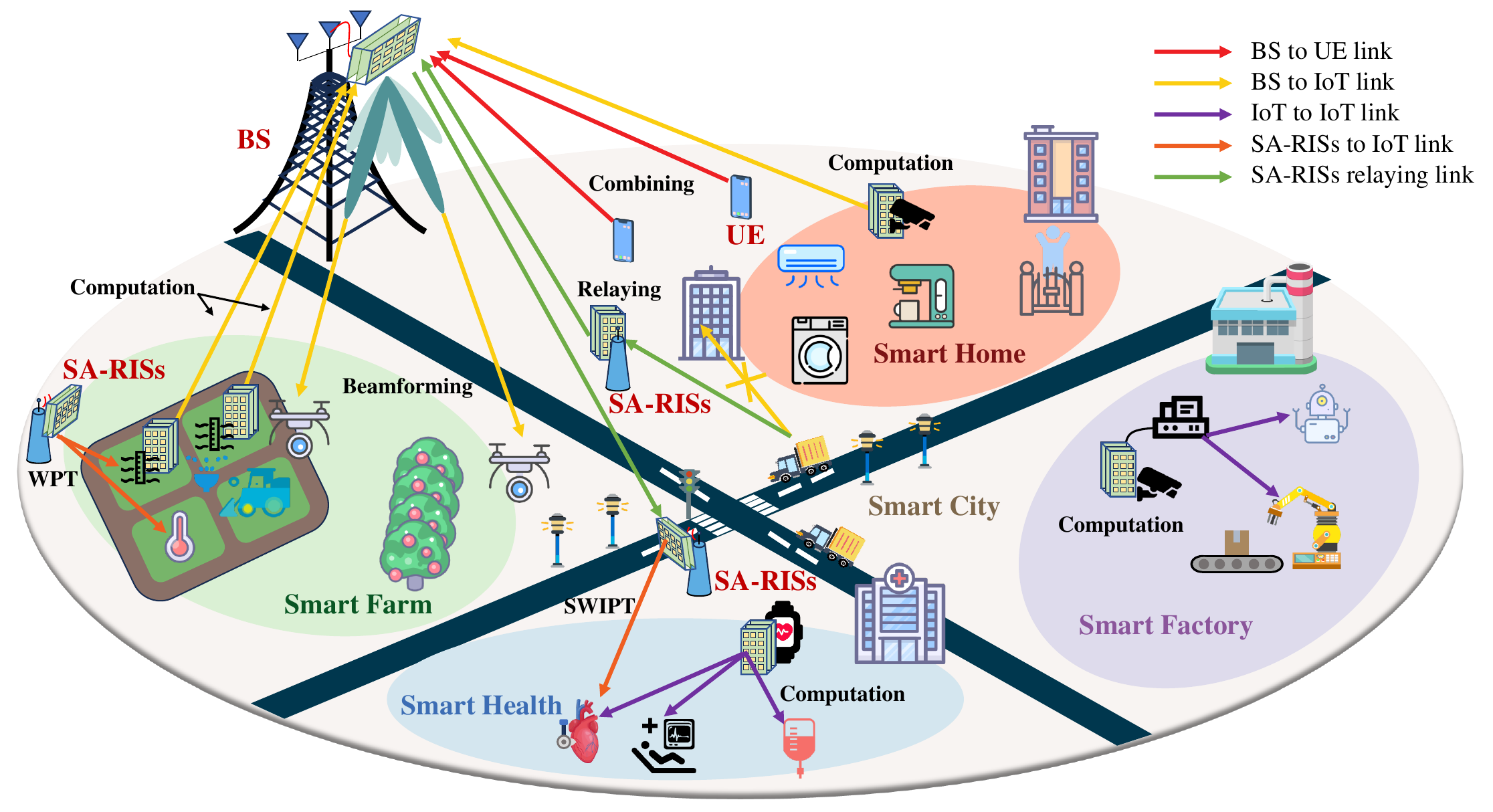}
\captionsetup{font={small}}
\caption{System architecture of multi-layer RIS-empowered edge IoT scenario, where multi-layer RIS is leveraged to facilitate the communications, computation, and long-term operation of edge IoT devices. This enables various applications such as smart cities, health, farms, homes, and factories, contributing to a smarter and more connected world.}
\label{system}

\end{figure*}

\section{System Architecture}
The system architecture of multi-layer RIS-empowered edge IoT scenario is illustrated in Fig.~\ref{system}. The coexistence of user equipments (UEs) and numerous IoT devices at the network edge reflects the complexity and diversity of future networks, highlighting the challenges of managing on-edge services. Multi-layer RIS can be configured on the BS and IoT devices, or operate independently, i.e., standalone multi-layer RIS (SA-RIS). Under different configurations, multi-layer RIS can achieve MIMO communications, perform computation, and enable WPT thanks to its enhanced computation capability. The universal structure with versatile functions makes multi-layer RIS essential for empowering future networks. The major data and energy transmission links in the considered scenario, along with the specific functions of multi-layer RIS, are detailed below.
\begin{itemize}
    \item\textbf{BS to UE links.} Enable UEs to access multimedia services through the network. Multi-layer RIS is deployed at the BS to perform transmit beamforming or receive combining directly in the EM wave domain.
    \item\textbf{BS to IoT links.} Enable uplink data uploading and downlink control signal transmission between IoT devices and the BS. Multi-layer RIS can be configured on the BS and IoT devices to perform certain signal processing tasks to facilitate transmission, such as data compression and coding. Additionally, some computation tasks can be executed by multi-layer RIS to facilitate information extraction from raw data. Downlink control signal transmission to multiple IoT devices can be achieved through beamforming executed by multi-layer RIS on the BS.
    \item\textbf{IoT to IoT links.} Support collaboration and autonomous decision-making among IoT devices. Multi-layer RIS provides  computation capability to IoT devices.
    \item\textbf{SA-RIS to IoT links.} Realize energy transfer to IoT devices. An SA-RIS structure acts as a power beacon to power IoT devices via WPT. Besides, the SA-RIS can leverage SWIPT to power IoT devices while relaying control signals from the BS to them. Energy harvested by IoT devices enables wireless information transfer (WIT) in the next phase.
    \item\textbf{SA-RIS relaying links.} Reconfigure the wireless propagation environment by overcoming the physical blockage to facilitate data transmission among BS, UEs, and IoT devices. In this situation, an SA-RIS act as a relay.
\end{itemize}

Notably, the introduction of multi-layer RIS requires minimal changes to the traditional communication architecture. Specifically, taking the transmitter as an example, only the transmission process of radio frequency (RF) signals is altered, as they pass through multi-layer RIS before being emitted to the wireless channel. No modifications are required in the baseband unit, except that some baseband signal processing tasks can be offloaded to multi-layer RIS for execution in the wave domain. In addition to multi-layer RIS itself, the only supplementary hardware required is an FPGA controller. Due to the nearly passive and low-cost nature of meta-atoms, along with the simplified hardware architecture, multi-layer RIS offers a highly energy-efficient and economically viable solution. In conclusion, deploying multi-layer RIS in reality offers significant advantages, including low hardware complexity, reduced cost, minimal power consumption, and backward compatible with existing IoT devices and communication infrastructure.

Despite the minor change in hardware architecture, improved performance in MIMO transmission and WPT can be achieved due to the more precise beam-steering capabilities of multi-layer RIS. Based on this, multi-layer RIS introduces computation capability, enabling a universal paradigm of integrated communication and computation. When executing simple computation tasks,  multi-layer RIS-based schemes offer significant advantages over traditional approaches, such as light-speed, power-free computation,  and the ability to handle multiple tasks simultaneously. These advantages make multi-layer RIS well-suited for future IoT and edge computing scenarios, where vast amounts of data need to be processed in a timely and energy-efficient manner. 

To clearly demonstrate the capabilities of multi-layer RIS, the following section elaborates on multi-layer RIS-enabled MIMO transmission, computation, and WPT schemes in sequence. The performance analysis, potential applications, and advantages of each multi-layer RIS-based scheme are discussed in detail, with a focus on IoT scenarios.

\section{Multi-layer RIS on Edge}
\subsection{Multi-layer RIS-enabled communication}
As IoT device density increases, the current deployment of MIMO antennas at BS struggles to mitigate inter-signal interference while serving numerous devices simultaneously. Nevertheless, adding more antennas to mitigate interference introduces further challenges, including higher hardware cost, increased power consumption, and greater algorithmic complexity. To overcome these limitations, multi-layer RIS can be configured at the BS to replace phased arrays or even all beamforming components. By dynamically adjusting its EM response, multi-layer RIS allows the information-bearing EM waves to perform beamforming at the speed of light in the wave domain during forward propagation. Therefore, the processing time and computational complexity are independent of the dimension of matrix and the number of mathematical operations, resulting in better scalability of antenna units and serving devices. Besides, this novel transmitter paradigm supports MIMO communication with reduced hardware cost and complexity by eliminating the need for excessive RF chains and high-precision DACs.

Besides, integrating large-scale MIMO antennas into small-sized, energy-constrained IoT devices to facilitate data transmission remains challenging. Multi-layer RIS addresses this challenge by tightly stacking multiple RIS layers with densely packed meta-atoms, enhancing data transmission rates while reducing the physical dimensions of devices. Additionally, wave-based beamforming does not incur extra power consumption compared to conventional RF signal
processing, which is more conducive to ensuring the long-term operation of IoT devices.

In summary, compared to traditional digital  MIMO architectures, the multi-layer RIS-based scheme provides an energy-efficient, cost-effective, and compact solution for existing IoT scenarios.

\begin{figure}
\centering
\includegraphics[width=8.5cm]{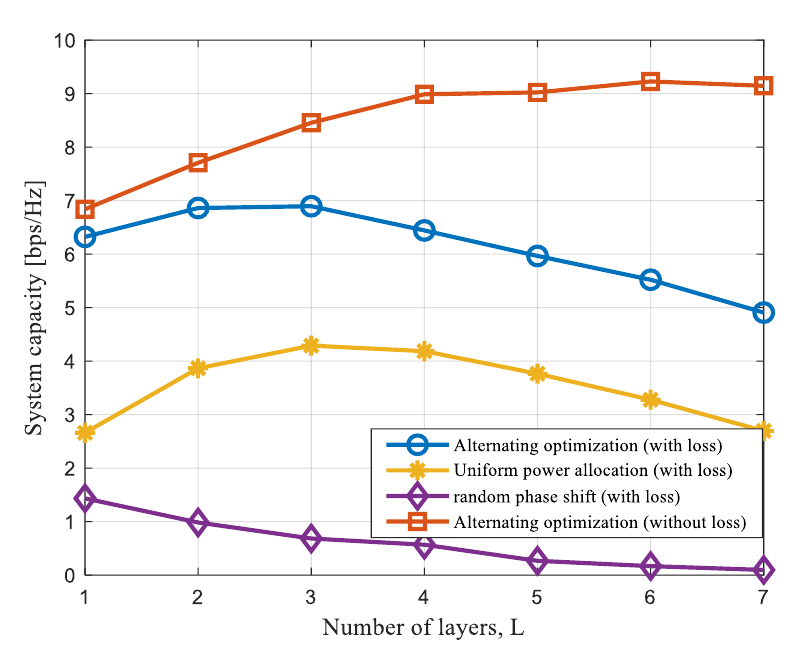}
\captionsetup{font={small}}
\caption{System capacity of multi-layer RIS-enabled MIMO transmission with respect to the number of RIS layers.}
\label{SE}

\end{figure}

To better demonstrate the capability of the multi-layer RIS-based MIMO transmission scheme in improving system capacity, a simple example is investigated. In the setup, a BS simultaneously serves four devices with the assistance of multi-layer RIS, under a maximum transmit power budget. Assuming the phase shifts of multi-layer RIS can be continuously tuned, an alternating optimization algorithm is employed to maximize system capacity. The algorithm includes an iterative water-filling  algorithm for transmit power allocation and a projected gradient ascent algorithm for phase shift optimization~\cite{alg}. Fig.~\ref{SE} illustrates the system capacity versus the number of  RIS layers. In the ideal case, benefiting from the capability of multi-layer RIS to suppress the inter-signal interference, the system capacity increases with the number of layers before gradually saturating. However, the penetration losses incurred as signals pass through each layer must also be considered. In this case, when the number of RIS layers becomes large, the gain introduced by multi-layer RIS is insufficient to compensate for the penetration losses. The optimal number of layers identified in this study is 3. Furthermore, the alternating optimization algorithm outperforms both the random phase shift and uniform power allocation schemes, demonstrating its effectiveness. 

\subsection{Multi-layer RIS-enabled computation}

\begin{figure*}
\centering
\includegraphics[width=16.5cm]{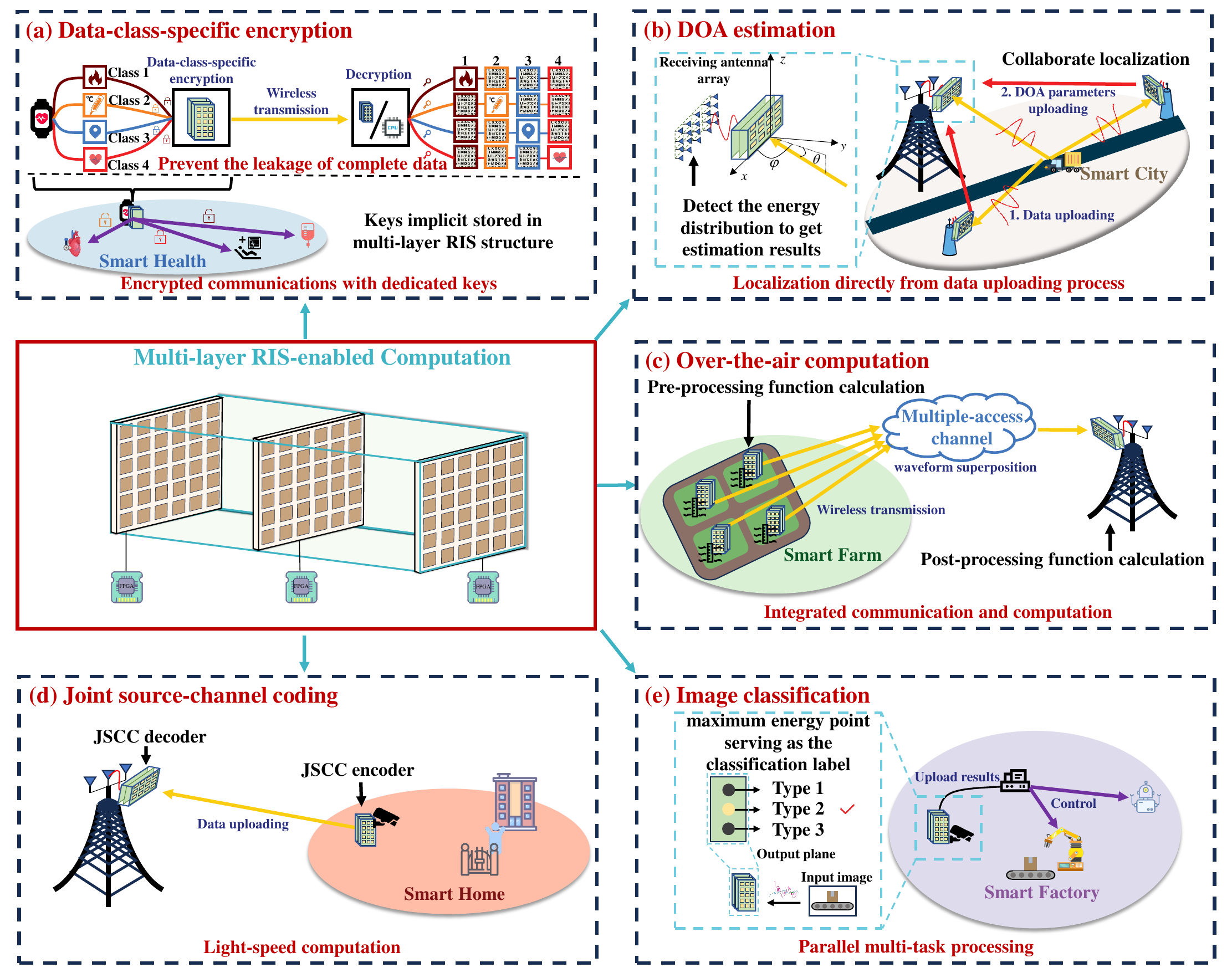}
\captionsetup{font={small}}
\caption{Multi-layer RIS-enabled computation schemes, where multi-layer RIS is utilized to perform (a) data-class-specific encryption; (b) DOA estimation; (c) over-the-air computation; (d) joint source-channel coding; (e) image classification.}
\label{computation}

\end{figure*}

As shown in Fig.~\ref{computation}, various computation tasks can be efficiently executed using multi-layer RIS, enabling numerous smart applications in edge IoT scenarios. Specifically, these tasks include signal processing for wireless communications and AI-related functions. The principles, advantages, and potential applications of each multi-layer RIS-based computation scheme are detailed below.

A $\rm D^2NN$-based data-class-specific encryption scheme was proposed in~\cite{coding}. The  trained $\rm D^2NN$ can differentiate data from different classes and perform class-specific encryption using unique linear transformation matrices (encryption keys) preassigned to each data class. On the receiving side, only the correct inverse matrix (decryption key) can successfully recover the original information. This method enhances privacy protection by preventing the complete exposure of data. For instance, class-specific encryption can be applied to heterogeneous data (e.g., temperature, blood pressure, and user location) collected by multimodal sensors in smart health applications. In this case, the leakage of one key does not result in the exposure of all data. What’s more, distributing unique keys to each IoT device in communication is unrealistic, as IoT devices with limited storage space would be required to  store all the keys of the devices they interact with. This scheme solves the issue by making the key implicitly stored in the $\rm D^2NN$ structure and executing encryption as the EM wave passes through multi-layer RIS.

A multi-layer RIS-based direction-of-arrival (DOA) estimation scheme was proposed in~\cite{DOA}. After passing through the multi-layer RIS, a signal's direction can be estimated by detecting the energy distribution across the receiving antenna array. This approach allows DOA estimation to be performed by directly observing the spatial-domain spectrum, offering an energy-efficient and hardware-cost-effective solution. Many wireless sensor network applications, such as logistics management and asset management in smart cities, require integrating location information into IoT data. By combining the DOA estimation results and geometric relationships, the BS and SA-RISs can collaborate to localize IoT devices based on their uploaded data. This process significantly reduces the amount of transmitted data, as no additional location information needs to be included.

We propose a multi-layer RIS-based over-the-air computation scheme to facilitate data analysis at the edge server. Specifically, $\rm D^2NN$ has been shown to perform universal linear transformation~\cite{coding} and hold the potential to implement nonlinear activation functions, making it a universal function approximator. Leveraging these properties, multi-layer RIS can be utilized to perform pre-processing function calculations on local data at IoT devices and post-processing function calculations on the superposed data received at the edge server. This approach facilitates the implementation of a low-power over-the-air computation scheme, which is particularly beneficial for analyzing data in environments with large-scale IoT deployments. In smart agriculture, for instance, this scheme enables the computation of key functions that support decision-making. Notably, multiple functions can be computed simultaneously without inter-function interference by a multi-layer RIS-based beamformer during function calculations, thereby enabling a universal paradigm of integrated communication and computation. 

A multi-layer RIS-based over-the-air joint source-channel coding (AirJSCC) scheme was introduced in~\cite{JSCC}, transforming the computation process inherent in JSCC into the transmission of wireless signals through multi-layer RIS. This approach can significantly reduce data redundancy during transmission without compromising reliability through the joint optimization of source and channel coding, thereby offering superior performance in bandwidth-limited and energy-constrained IoT scenarios. It is noteworthy that computation is performed automatically at each layer as the EM wave propagates through multi-layer RIS, known as light-speed computation. These advantages make the scheme ideal for real-time monitoring applications, such as child care and anti-theft surveillance in smart homes, where large volumes of data must be uploaded to cloud servers in a timely manner to ensure continuous remote monitoring.

A $\rm D^2NN$-based handwritten digit classification scheme was constructed in the microwave frequency in~\cite{classification}. Specifically, the excited EM wave first passes through digit patterns to be classified, completing the modulation process. The modulated EM wave then propagates through the $\rm D^2NN$ for computation. Upon exiting the $\rm D^2NN$, the wave focuses on designated focal points on the output plane, with the point of maximum energy determining the classification label for direct digit recognition. By adapting the training set, multi-layer RIS-based classification scheme can be extended to handle other types of classification tasks. Notably, multiple tasks can be processed simultaneously by encoding them into distinct polarization channels~\cite{polar}. This unique feature makes the scheme particularly well-suited for real-time multi-task scenarios in industrial IoT, such as product classification and quality inspection, where automated sorting and defective product rejection can be done simultaneously.

Since multi-layer RIS is a physical realization of neural networks, a similar conclusion about computation performance can be drawn. As the number of RIS layers increases, the computation performance of multi-layer RIS-enabled schemes improves accordingly, until the vanishing gradient problem becomes more pronounced, making further improvement of the neural network difficult. Interested readers can refer to~\cite{shortcut} and~\cite{DOA} for simulation demonstrations of this conclusion.

In summary, when compared to traditional hardware-based computation approaches, multi-layer RIS-based schemes offer a solution with  low computational cost. Specifically, thanks to the nearly passive nature of the meta-atoms, the power consumption in multi-layer RIS-based computation schemes is primarily attributed to the FPGA used for system control, which is significantly lower compared to traditional digital hardware (such as graphics cards)~\cite{semantic}. Beside, as computation is performed in the wave domain, multi-layer RIS achieves low latency that is independent of the number of neurons (meta-atoms) and the input data dimension. Therefore, in addition to its relatively low hardware cost, multi-layer RIS offers a highly cost-effective computation solution.

\subsection{Multi-layer RIS-enabled WPT}
Since WPT aims to enhance energy transfer efficiency, its integration with low-power RIS to reduce hardware power consumption is inherently advantageous. With the miniaturization of IoT devices, improving the resolution of energy beamforming to achieve effective differentiation and accurate energy transfer has become a key challenge. However, the application of single-layer RIS in WPT encounters unavoidable challenges in improving beam resolution. According to~\cite{compact}, when a single-layer RIS is used for beamforming, most of the transmission power is concentrated in the central meta-atoms, while the non-central ones contribute minimally. In contrast, with multi-layer RIS, the power distribution becomes more balanced in the second layer through phase shift control in the first layer. This phenomenon indicates that more meta-atoms in multi-layer RIS contribute to beamforming, resulting in higher beam resolution compared to its single-layer counterpart, thereby improving energy transfer accuracy and efficiency.

Besides, compared to BS-based WPT schemes, SA-RIS-based solution offers flexible deployment closer to IoT devices, improving energy transfer efficiency regardless of device location. This ensures that IoT devices can operate for extended periods without the need for battery replacement. SWIPT can also be leveraged by the SA-RIS to relay necessary control signals and power IoT devices simultaneously. What’s more, the reduced distance between the SA-RIS and IoT devices also allows for the exploitation of near-field communication. Specifically, as higher frequency bands and larger apertures of RIS are used to meet the higher demands of future networks, near-field effects become increasingly significant. 
In near-field channels, energy beamfocusing can be concentrated on specific locations determined by both direction and distance, resulting in better differentiation and higher energy concentration at IoT devices.

\begin{figure}
\centering
\includegraphics[width=9cm]{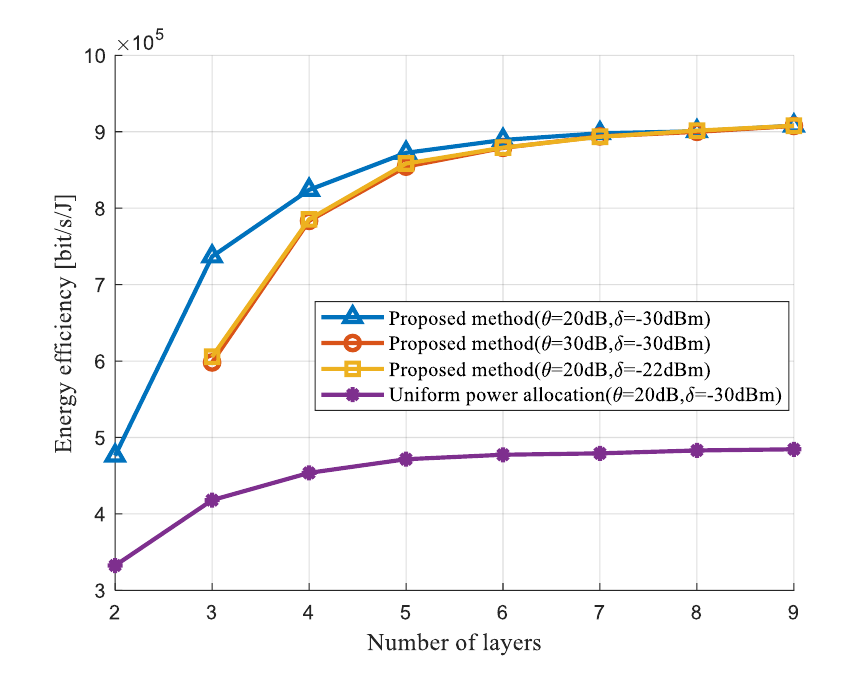}
\captionsetup{font={small}}
\caption{Energy efficiency of the multi-layer RIS-enabled SWIPT scheme with respect to the number of RIS layers, where $\theta$ is the minimum SINR requirement and $\delta$ is the minimum energy harvest requirement.}
\label{EE}

\end{figure}

A case study on SWIPT is presented to demonstrate the potential of multi-layer RIS in enhancing energy efficiency. In this scenario, an SA-RIS are used to power four IoT devices while simultaneously transmitting control signals. Two quality of service (QoS) targets are considered for  IoT devices to ensure  sustained operation and reliable control, i.e., minimum energy harvest requirement and minimum signal-to-interference-plus-noise ratio (SINR) requirement. Given the limited power budget of the SA-RIS, an optimization problem is formulated to maximize system energy efficiency, subject to these QoS constraints. The simulation results, depicted in Fig.~\ref{EE}, show that system energy efficiency improves as the number of RIS layers increases until it eventually saturates. When comparing setups with different QoS targets, it is evident that better energy efficiency is achieved when the QoS targets are less stringent. In cases where stricter QoS targets are imposed (i.e., higher energy harvest or SINR requirement), a 2-layer structure fails to meet the QoS targets. However, due to the capability of multi-layer RIS, the performance gap between different QoS target setups gradually diminishes as the number of layers increases. This finding suggests that multi-layer RIS can improve signal processing capabilities by stacking additional layers, thereby increasing the degrees of freedom to better meet QoS targets and reduce performance disparity across different QoS configurations. Additionally, compared with the uniform power allocation scheme, the proposed method demonstrates its effectiveness.

\section{Future Directions}
Despite the aforementioned benefits of multi-layer RIS, several challenges remain to be addressed for its practical implementation. Key future research directions are outlined below.

\subsection{Inter-layer channel modeling}
The Rayleigh-Sommerfeld diffraction theory, widely used to model inter-layer channels, is based on the wave nature of light. It is derived from the Huygens-Fresnel principle and subsequently developed into a mathematical model. Due to inconsistencies in meta-atom fabrication and small deformations during the assembly process of multi-layer RIS, the actual inter-layer channel coefficients may deviate from the ideal Rayleigh-Sommerfeld diffraction equation. In the multi-layer RIS-based schemes proposed in this paper, the inter-layer channel is typically treated as a known quantity, and multi-layer RIS is optimized or trained accordingly. However, when the actual channel deviates from the theoretical value, beam pointing will be misaligned and the computation results will become inaccurate. Therefore, it is essential to calibrate the inter-layer channel coefficients before deploying multi-layer RIS. This calibration can be achieved, for example, by applying the classic error back-propagation algorithm to iteratively update the channel coefficients obtained through theoretical calculations, until the calculated output field closely matches the measured values. Still, further in-depth investigation and practical measurement campaigns are required to refine the inter-layer channel model, thereby facilitating the continued research and implementation of multi-layer RIS.

\subsection{Resource allocation and scheduling}
In the considered edge IoT scenario, both WPT and WIT are supported by multi-layer RIS. Therefore, it is crucial to develop scheduling strategies that minimize co-channel interference between these two processes. By employing synchronization techniques in the network, interference between the two stages can be canceled in the time domain. By applying successive interference cancellation techniques, deterministic WET signals can be eliminated, thereby enhancing the performance of WIT. Additionally, the transmit power of IoT devices for data transmission is dependent on the energy harvested from WPT, creating a coupling between WPT and WIT. For instance, when an IoT device needs to upload large amounts of data, more power and time slots should be allocated to the SA-RIS for WPT, with multi-layer RIS-based energy beamforming to enhance accuracy and efficiency. However, integrating beamforming to coupled WPT and WIT with time-frequency resource allocation presents challenges, particularly considering the dynamic nature of the channel and the large-scale configuration of IoT devices. To address this, advanced techniques such as deep reinforcement learning present promising alternatives to traditional optimization-theory-based algorithms. These approaches treat resource allocation and scheduling as decision processes, interacting with the dynamic channel environment to discover optimal strategies through exploration and exploitation. Thereby, deep reinforcement learning has the potential to address the challenges in resource allocation arising from rapidly changing channels, the coupling between WPT and WIT, and large-scale networks,  facilitating the practical deployment of multi-layer RIS.

\subsection{Channel estimation}
In the proposed multi-layer RIS-based MIMO transmission and WPT schemes, accurate channel state information (CSI) for each device is essential to ensure precise beamforming. However, obtaining CSI in multi-layer RIS-assisted systems presents a significant challenge, as multi-layer RIS operating without sensing modules cannot directly estimate the channel.  Moreover, the implementation of large-scale RIS increases the complexity of the channel estimation algorithm and the number of pilots required.  This challenge is further compounded by the growing number of users and IoT devices.  One potential solution is to leverage compressive sensing and deep learning methods to exploit channel sparsity in the beamspace.  Nevertheless, due to inter-layer channel deviations and hardware impairments, the estimated CSI is prone to errors.  Therefore, robust beamforming design, which accounts for the impact of CSI errors on beamforming performance, is an important area of research.

\subsection{Edge training}
When executing multi-layer RIS-based computation tasks, a training process similar to that of traditional hardware-based deep neural networks is essential. For example, a gradient descent algorithm is needed to iteratively update the EM response of each meta-atom until convergence is achieved. In the multi-layer RIS-based classification and DOA estimation schemes, the loss functions can be defined as the error between the actual energy distribution on the output plane and the theoretical distribution determined by the ground truth. In the computation schemes with a dual-end multi-layer RIS configuration, such as JSCC, end-to-end information recovery error minimization is used to train the two multi-layer RIS architectures jointly. 
However, considering that multi-layer RIS is deployed in the edge IoT scenario, a distributed edge training procedure should be adopted to avoid the transmission of large volumes of data to a central data center and better handle the exponential growth of IoT devices. Federated Learning offers attractive benefits for the proposed multi-layer RIS-enabled smart applications. By training multi-layer RIS using local data collected by IoT devices and uploading only local updates, data privacy is ensured, which is crucial for smart factory and smart home applications. Additionally, by downloading a powerful global model constructed from numerous local models, the scalability and generalization ability of multi-layer RIS-based computation schemes are enhanced.

\section{Conclusions}
In this article, we have proposed a multi-layer RIS-based universal paradigm that enables MIMO communication, computation, and WPT simultaneously to support IoT devices at the network edge. The multi-layer RIS structure, similar to a neural network, transforms signal processing or computation into the process of wireless signals passing through multi-layer RIS. This novel approach offers unique advantages and innovative applications across these three functional domains, as discussed. Compared to traditional digital hardware methods, the multi-layer RIS-based approach demonstrates the potential for scalability at a low cost, facilitating light speed, power-free computation, and precise power transfer. Our simulations of multi-layer RIS-based MIMO transmission and SWIPT highlight enhanced system performance compared to the single-layer counterparts. In a nutshell, multi-layer RIS marks a significant paradigm shift from conventional digital hardware-based to wave-based computation, promising more energy-efficient wireless networks in the future.

\newpage
\balance

\section*{Biographies}

\vspace{-14 mm}
\begin{IEEEbiographynophoto}{Shuyi Chen}
(chenshuyitina@gmail.com) received her B.Sc, M.Sc and Ph. D. degrees in information and communication engineering from School of Electronics and Information Engineering, Harbin Institute of Technology (HIT), Harbin, China, in 2013, 2015 and 2021, respectively. She is currently an associate professor at HIT. Her research interests include performance analysis and interference avoidance in ultra-dense networks, as well as multi-layer reconfigurable intelligent surfaces (RIS).
\end{IEEEbiographynophoto}

\vspace{-14 mm} 
\begin{IEEEbiographynophoto}{Junhong Jia}
(24S005051@stu.hit.edu.cn) received his B.Sc. degree from the School of Information Science and Engineering, Harbin Institute of Technology (HIT), Weihai, China, in 2024. He is currently pursuing his  M.Sc. degree at the Harbin Institute of Technology (HIT), Harbin, China. His current research interests include multi-layer reconfigurable intelligent surfaces (RIS), and ultra-dense networks.
\end{IEEEbiographynophoto}

\vspace{-14 mm} 
\begin{IEEEbiographynophoto}{Baoqing Zhang}
(bqzhang10@163.com) received Ph.D. degree in 2014 from University of
Science and Technology Beijing. He is currently a senior engineer at Beijing Institute of Electronic System Engineering, China. His main research interests include information processing.
\end{IEEEbiographynophoto}

\vspace{-14 mm} 
\begin{IEEEbiographynophoto}{Yingzhe Hui}
(yingzhe\_hui@stu.hit.edu.cn) received his B.Sc. degree from the School of Computer and Information, Hefei University of Technology (HFUT), Hefei, China, in 2022. Currently, he is pursuing his M.Sc. degree at the Harbin Institute of Technology (HIT), Harbin, China. His current research interests include multi-layer reconfigurable intelligent surfaces (RIS), massive MIMO, and semantic communications.
\end{IEEEbiographynophoto}

\vspace{-14 mm} 
\begin{IEEEbiographynophoto}{Yifan Qin}
(qinyifan@hrbeu.edu.cn) received his B.S., M.S., and Ph.D. degrees from Harbin Institute of Technology, Harbin, China, in 2013, 2015, and 2022, respectively. He is currently with the Key Laboratory of In-Fiber Integrated Optics, Ministry of Education, Harbin Engineering University, and the Key Laboratory of Photonic Materials and Device Physics for Oceanic Applications, Ministry of Industry and Information Technology of China, Harbin Engineering University. His research interests include optical microscopy, fiber optical tweezers, intelligent fiber devices, and fiber sensors. He is the author and co-author of over 30 research papers.
\end{IEEEbiographynophoto}

\vspace{-14 mm} 
\begin{IEEEbiographynophoto}{Weixiao Meng}
[M'04, SM'10] (wxmeng@hit.edu.cn) received his B.E., M.E., and Ph.D. degrees from the Harbin Institute of Technology (HIT), Harbin, China, in 1990, 1995, and 2000, respectively. He worked with NTT DoCoMo as a Visiting Researcher from 1998 to 1999, and with the University of California, Riverside, as a Senior Visiting Scholar in 2017. Currently, he is a Full Professor with the School of Electronics and Information Engineering, HIT. He is a fellow of the China Institute of Electronics and a Senior Member of the IEEE ComSoc and the China Institute of Communications. He serves as the Chair of the IEEE ComSoc Harbin Chapter. He acted as a leading TPC Co-Chair of ChinaCom 2011 and ChinaCom 2016, Awards Co-Chair of ICC 2015 and Wireless Networking Symposia Co-Chair of Globecom 2015, AHSN Symposia Co-Chair of Globecom 2018 and ICC 2020, and leading Workshop Co-Chair of ICC 2019 and ICNC 2020.Under his leadership, the Harbin Chapter received the IEEE ComSoc Chapter of the Year Award, the Asia-Pacific Region Chapter Achievement Award, and the Personal Member \& Global Activities Contribution Award in 2018.
\end{IEEEbiographynophoto}

\vspace{-14 mm} 
\begin{IEEEbiographynophoto}{Tianheng Xu}
[M] (xuth@sari.ac.cn) received the Ph.D. degree from the Shanghai Institute of Microsystem and Information Technology, Chinese Academy of Sciences, Shanghai, China, in 2016. He is currently an Associate Professor with the Shanghai Advanced Research Institute, Chinese Academy of Sciences. His research interests include 5G/6G wireless communications, ubiquitous cognition and intelligent spectrum sensing technology, multimodal signal processing, and heterogeneous system interaction technologies. He received the President Scholarship of Chinese Academy of Sciences (First Grade), in 2014, the Outstanding Ph.D. Graduates Award of Shanghai, in 2016, best paper awards at the IEEE GLOBECOM 2016 and the Springer MONAMI 2021, the First Prize of Technological Invention Award from the China Institute of Communication, in 2019, and the First Prize of Shanghai Technological Invention Award, in 2020.
\end{IEEEbiographynophoto}

\vfill

\end{document}